\documentclass[prl, reprint]{revtex4-1}
\usepackage{natbib}
\usepackage{graphicx}

\begin{document}

\title{Persistent narrowing of nuclear-spin fluctuations in InAs quantum dots using laser excitation}
\date{\today}
\author{Bo Sun}
\author{Colin Ming Earn Chow}
\author{Duncan G. Steel}
\email{dst@eecs.umich.edu}
\affiliation{The H. M. Randall Laboratory of Physics, The University of Michigan, Ann Arbor, Michigan 48109, USA}
\author{Allan S. Bracker}
\author{Daniel Gammon}
\affiliation{Naval Research Laboratory, Washington D.C. 20375, USA}
\author{L. J. Sham}
\affiliation{Department of Physics, The University of California, San Diego, La Jolla, California 92093, USA}


\begin{abstract}
We demonstrate the suppression of nuclear spin fluctuations in an InAs quantum dot and measure the timescales of the spin narrowing effect. By initializing for tens of milliseconds with two continuous wave diode lasers, fluctuations of the nuclear spins are suppressed via the hole assisted dynamic nuclear polarization feedback mechanism. The fluctuation narrowed state persists in the dark (absent light illumination) for well over one second even in the presence of a varying electron charge and spin polarization. Enhancement of the electron spin coherence time (T2*) is directly measured using coherent dark state spectroscopy. By separating the calming of the nuclear spins in time from the spin qubit operations, this method is much simpler than the spin echo coherence recovery or dynamic decoupling schemes.
\end{abstract}

\maketitle

Single electrons trapped inside self assembled quantum dots form a well defined and optically accessible qubit, and are featured as the central element of many proposed quantum logic devices\cite{xu2008,press2008,kim2010,kim2008,kim2011,greilich2009,carter2010}. However, the electron spin in a quantum dot is coupled to many ($10^4 - 10^5$) nuclear spins, primarily via the Fermi contact hyperfine interaction, whose fluctuations form the main contribution to electron spin dephasing at cryogenic temperatures\cite{merkulov2002,braun2005,yao2006,cywiski2009}. Thus, there has been considerable interest in suppressing electron spin dephasing by manipulating the nuclear spin ensemble\cite{stepanenko2006,giedke2006,klauser2006,rudner2007,danon2008,vink2009,latta2009,ladd2010,bluhm2010,coish2010,yang2010}, and recent results have shown that it is possible to protect the coherence of an ensemble of dots \cite{greilich2006,greilich2007} and even recover the electron spin coherence in a single quantum dot \cite{press2010}.

In this Letter, we use the hole assisted nuclear feedback mechanism\cite{xu2009,yang2010} to demonstrate the reproducible preparation of the nuclear magnetic field (Overhauser field) to a fluctuation suppressed state, considerably enhancing the electron spin coherence. This nuclear spin narrowing (NSN)\cite{stepanenko2006,giedke2006,klauser2006,rudner2007,danon2008,vink2009,latta2009,ladd2010,bluhm2010,coish2010,yang2010} is accomplished without the creation of large nuclear spin polarizations. The spin narrowed state can be prepared in tens of milliseconds, and persists for well over a second even in the presence of a fluctuating electron charge and spin\cite{maletinsky2007,maletinsky2009}. We directly measure the enhanced T2* using coherent dark state spectroscopy.

\begin{figure}

  \includegraphics[scale=.4]{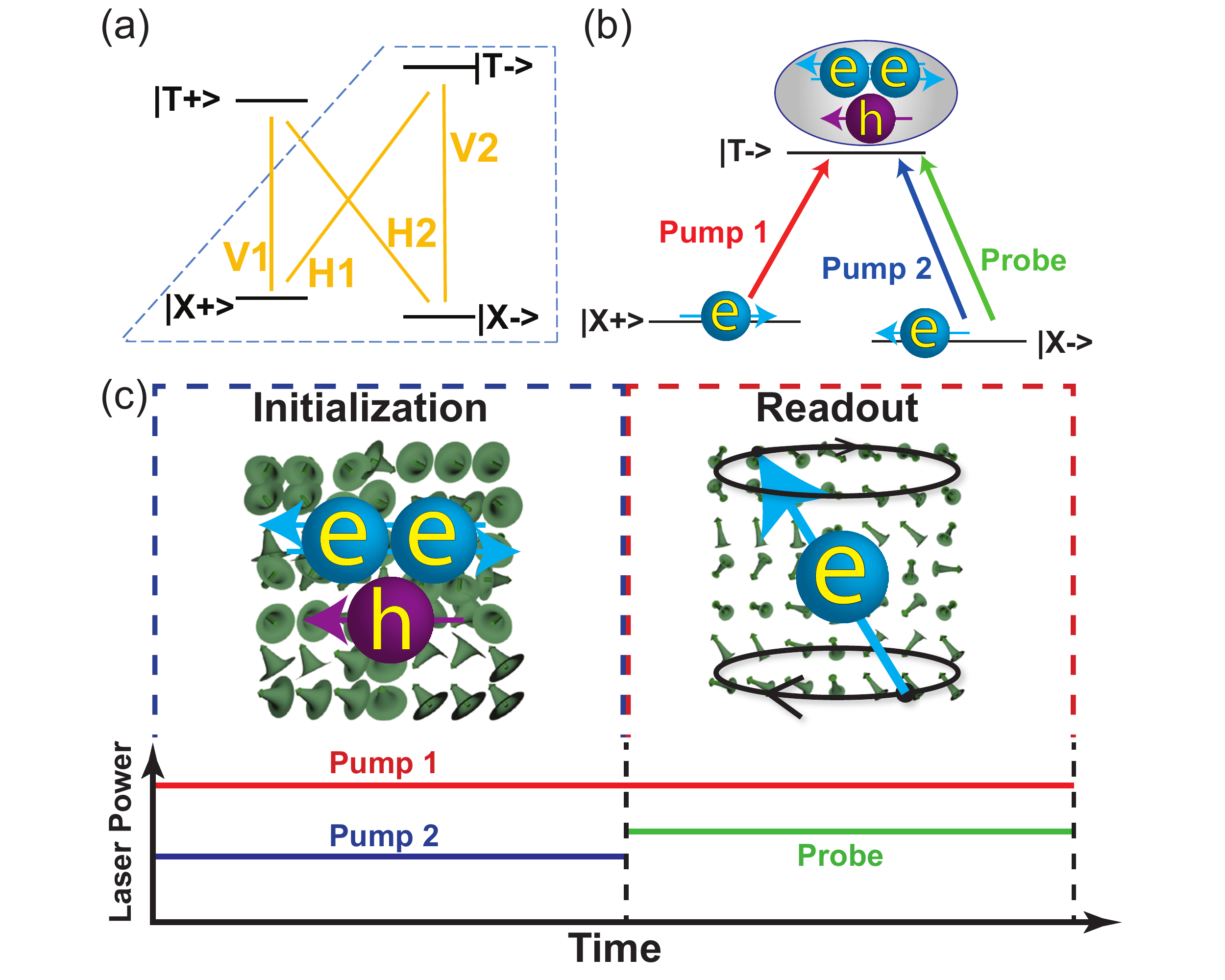}
  \caption{ (Color online). (a) Four level energy diagram for the trion system with a B=2.64T magnetic field applied in the Voigt profile. The $|T\pm>$ are the excited trion states and the $|X\pm>$ are the electron ground states. The relevant three level lambda sub-system is given by the dashed line. (b) Pump 1 is nearly resonant with transition H1, pump 2 and the probe are nearly resonant with transition V2. (c) Cartoon illustrating the laser illumination on the sample at each point in the scan. During the Initialization stage, pump 1 and pump 2 produce a trion, whose hole component interacts with the nuclear spins, preparing a NSN state. During the Read-out stage, pump 1 and the probe then produce and measure electron spin coherence, quantifying the narrowing of the nuclear spin distribution. The nuclear spins (green arrows in the background with large gaussian envelopes) start in a state of large fluctuation. The NSN state is represented by narrower gaussian envelopes, but maintains a similar average field.}\label{Fig1}
\end{figure}

The sample is an InAs self assembled quantum dot embedded in a Schottky diode structure. A DC bias voltage applied across the sample charges the dot with a single electron and Stark shift modulation spectroscopy a large amplitude AC component (0.08VAC) at 3.5Khz directly measure the absorption spectrum\cite{alen2003}. Applying a 2.64T magnetic field in the Voigt profile (perpendicular to the growth axis) turns on spin flip Raman transitions between the spin ground states ($|X\pm>$) and the charged exciton (trion) states ($|T\pm>$) (Fig. \ref{Fig1}a).
We selectively excite a three-level lambda ($\Lambda$) subsystem (dashed outline in Fig. \ref{Fig1}a) with narrow linewidth continuous wave lasers (Fig. \ref{Fig1}b). The lasers are passed through acousto-optical shutters to individually gate the lasers on and off, decoupling NSN initialization from electron spin control and readout. Pump 1 is resonant with the H1 transition while pump 2 is slightly detuned from the V2 transition. The probe scans across the V2 transition. Figure \ref{Fig1}c shows the gating of the lasers at each point of the absorption spectrum as the probe steps through the V2 resonance.

\begin{figure}

  \includegraphics[scale=.4]{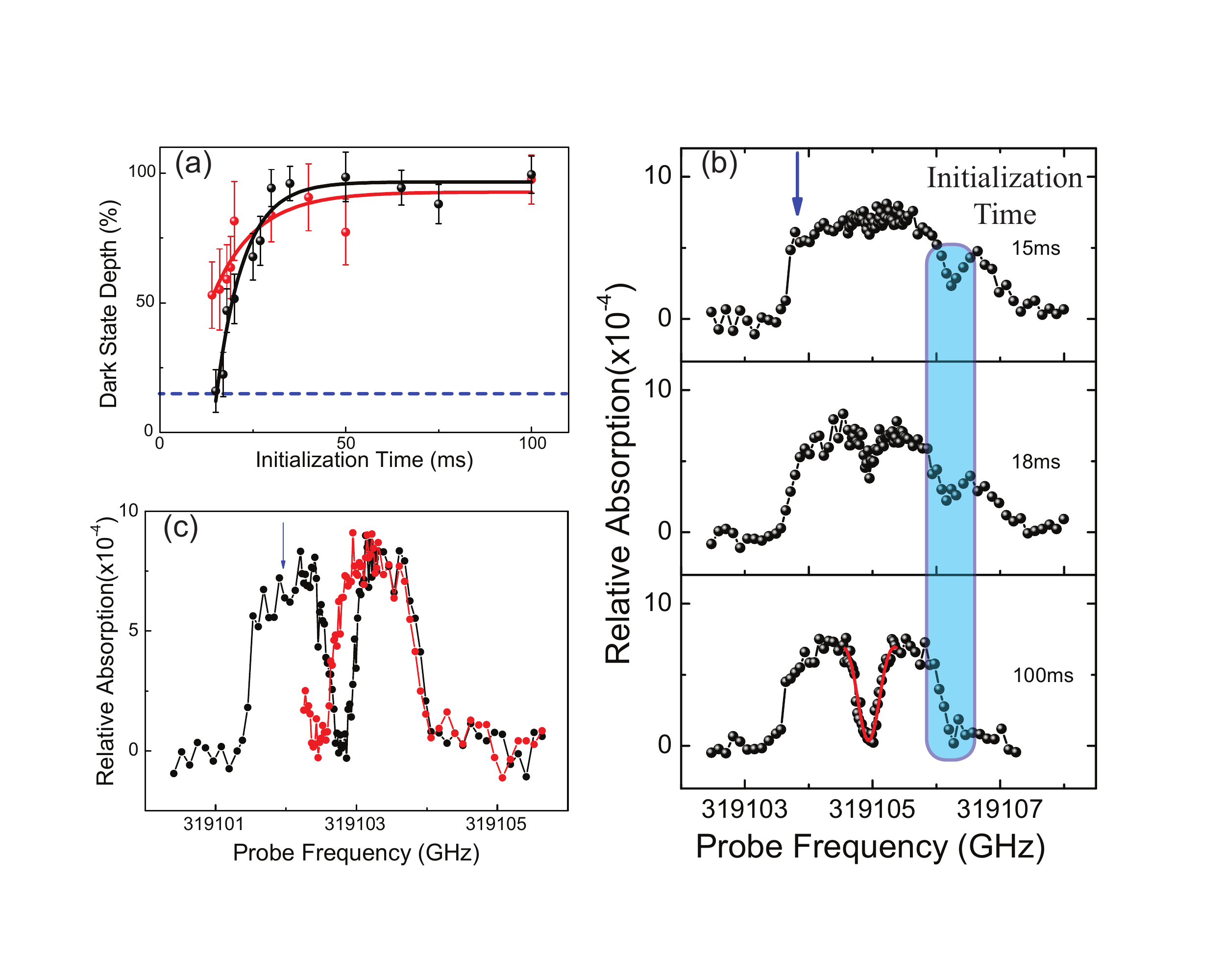}
  \caption{(Color online). (a) Measured dark state depths (relative to the Rabi sidebands) as a function of the initialization time at sample temperatures of 5K (black) and 14K (red). The lines are fits to exponentials from which we can extract a $1/e$ time of $7\pm 1$ms at 5K and $12\pm 6$ms at 14K. The dashed blue line indicates the relative depth of the dark state for expected thermal value of the electron spin decoherence rate of 360MHz. (b) Absorption spectra corresponding to select points in (a). At short initialization times a second dip appears to the blue (highlighted in blue), indicating bistability of the Overhauser field. The red solid line is a fit to the optical Bloch equations. (c) The black spectrum is taken using a nominal scan range. The red spectrum has a reduced scan range and shows a corresponding shift of the TPR, due to probe effects on the Overhauser field. The blue arrows indicate the location of pump 2. }\label{Fig2}
\end{figure}

To measure the onset time of the spin narrowing effect, pump 1 and pump 2 are first gated on, populating the trion state. The trion's unpaired hole interacts with the nuclei via a non-collinear hyperfine coupling, locking the Overhauser field to a value determined by the laser frequencies and produces spin narrowing via an intrinsic dynamic nuclear polarization feedback process\cite{xu2009}. Next, pump 2 is gated off and the probe is gated on for 25ms. When the pump and probe laser detunings are equal (at the two photon resonance), the electron spin forms a coherent superposition (dark state) which appears as a dip in the probe absorption spectrum\cite{gray1978,harris1997}. The strength of the dip is proportional to the electron spin decoherence rate (1/T2*)\cite{xu2008}. Since nuclear spin fluctuations contribute significantly to T2*, dark state spectroscopy is a sensitive measure of spin narrowing\cite{stepanenko2006,xu2009}. Data is read through a lock in amplifier with the integration time constant set to a small value (5ms) to minimize the readout time and thus minimize any perturbations to the nuclei due to the readout. We only integrate the signal during the read out phase to maintain a large signal strength. Figure \ref{Fig1}c shows the laser gate timing diagram for each point in a probe absorption spectrum.

Figure \ref{Fig2}a shows, for various initialization times, the measured probe absorption inside the dark state dip normalized to the absorption at the Rabi sidebands at sample temperatures of 5K (black curve) and 14K (red curve) and pump 1 (pump 2) Rabi frequency ($\Omega_R/2 \pi$) of 700MHz(150MHz). Fitting the data to an exponential (solid lines), we extract an $1/e$ of the NSN onset time of $7\pm 1$ms at 5K and $12\pm 6$ms at 14K. At the higher temperature, we expect an increase in hole relaxation to weaken the locking effect, requiring more time to generate the NSN state.  However, there is an almost factor of 2 decrease in the absolute signal strength, resulting in larger measurement noise.  Hence, the data in \ref{Fig2}a are not adequate evidence for this claim.

Solving the optical Bloch equations\cite{scully1997,meystre2007}, $i \hbar \frac{d \hat{\rho}}{d t}  = ([\hat{H},\hat{\rho} ] + \mathrm{decay})$ where $\hat{H}$ is the Hamiltonian and $\hat{\rho}$ is the density matrix, for a strong pump and weak probe in the lambda system, we can find the probe absorption at the dark state dip ($\alpha_{dip}$) and the Rabi sideband ($\alpha_{peak}$)\cite{xu2009}

\begin{equation}
\alpha _{dip}= \alpha _0 \frac{\chi ^2 \gamma _s+\gamma _t \left(\gamma _s{}^2\right)}{\chi ^4+ 2 \chi ^2\gamma _t\gamma _s+\gamma _t{}^2\gamma _s{}^2}
\end{equation}
\begin{equation}
\alpha _{peak}= \alpha _0 \frac{\chi ^2 \gamma _s+\gamma _t \left(\gamma _s{}^2+\chi ^2\right)}{ 2 \chi ^2\gamma _t\gamma _s+\gamma _t{}^2\gamma _s{}^2+\left(\gamma _t{}^2+\gamma _s{}^2\right)\chi ^2}
\end{equation}
where $\chi$ is half the pump 1 Rabi frequency, $\gamma_t$ is the trion dephasing rate, $\gamma_s$ is the electron spin dephasing rate, and  $\alpha_0$ is a constant. In the limit where $\gamma_s\ll \chi , \gamma_t$, the ratio between the dip and peak absorption reduces to  $\frac{\alpha _{dip}}{\alpha _{peak}} \approx \frac{\gamma _t \gamma _s}{\chi ^2 }$. Using this method, we estimate that $\gamma_s/2 \pi$=2MHz for a 100ms initialization time (at 5K). We also fit the dark state portion of the spectrum (solid red line in Fig. \ref{Fig2}b) with the optical Bloch equations (the spectrum is too distorted to fit directly) and find $\gamma_s/2\pi$=6MHz with an upper bound error of 14MHz. The expected thermal value of $\gamma_s/2\pi$ is 360MHz\cite{xu2009} at 5K and the relative dark state depth for this, calculated from simulations, is shown as the dashed blue line in Fig. \ref{Fig2}a.

Figure \ref{Fig2}b shows example spectra taken at 5K for several initialization times, where the blue arrow represents the position of pump 2 and each figure is an average of 40 scans. At short initialization times, there is a probe induced buildup of the Overhauser field which pushes the dark state to higher energy(highlighted in blue)\cite{xu2009}. At long initialization times, the probe effect is minimal. At intermediate initialization times, the appearance of the second dip may contribute to weakening of the central dip.

We note that while the dark state dip does not depend on the number of averages, which would be expected if there was an accumulated effect, we cannot discount memory effects in the dot which may impact the NSN dynamics measurement. The exponential function we use in Fig. \ref{Fig2}a is only intended to give an indication of behavior and is not meant to represent a physical model.

Because the measured $1/e$ times are less than the read out time, the Overhauser field weakly locks to the probe laser as it steps through the absorption spectrum\cite{xu2009}. This locking only occurs over a limited range and results in the distorted lineshapes of Fig. \ref{Fig2}b and c and shifts the TPR with a change in start position of the laser scan, seen in the red curve in Fig. \ref{Fig2}c compared to the black curve. This does not affect the NSN measurement, only the average Overhauser field build up as there is no hole population at the TPR. Hence, the influence of the finite readout time does not impact our conclusion regarding the time scale of the preparation of the NSN state.

\begin{figure}

  \includegraphics[scale=.4]{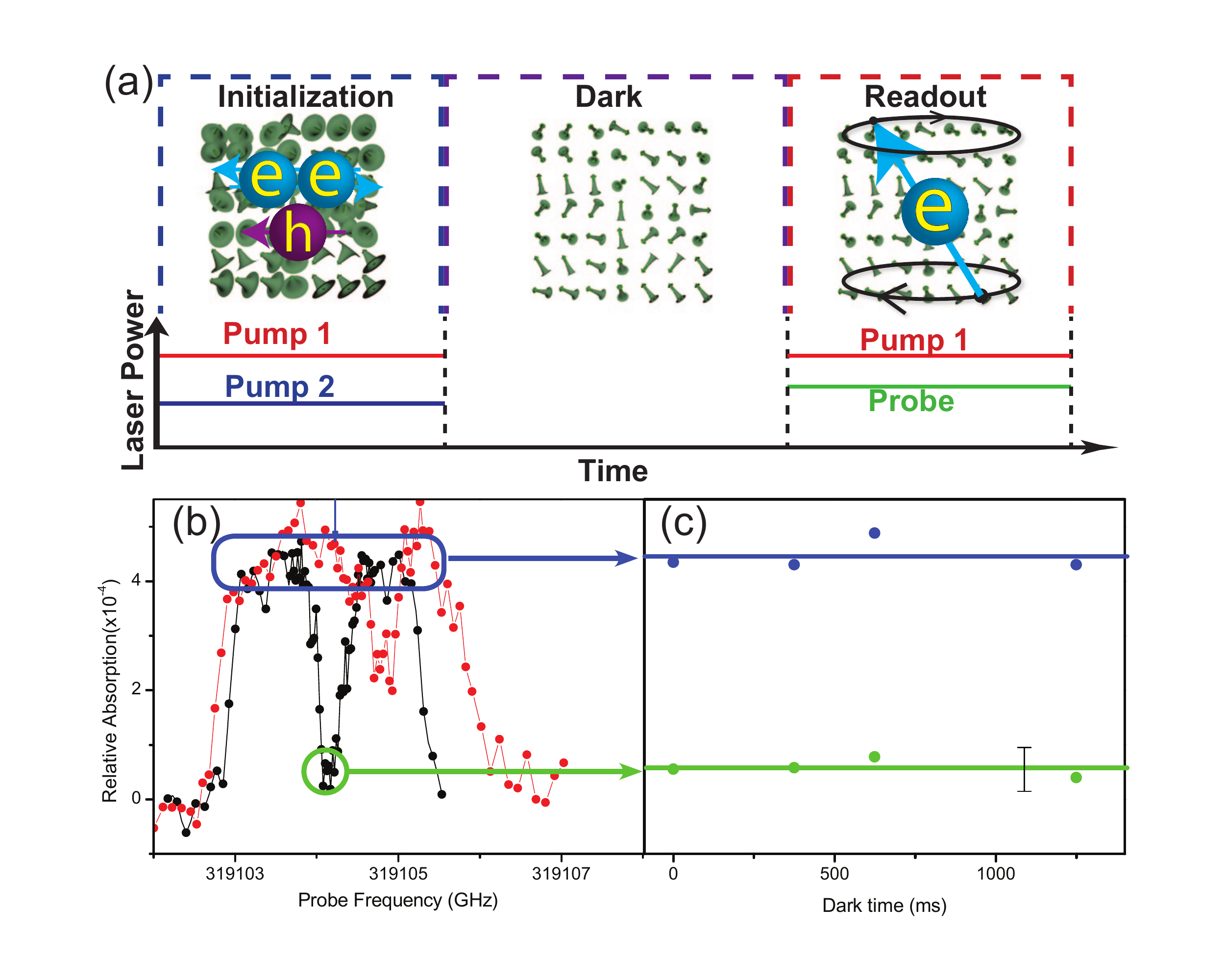}
  \caption{(Color online). (a) We insert a dark period into the gating sequence to measure the persistence of the NSN as a function of laser dark time. (b) Black data is the absorption spectrum for 0ms dark time. The red data is a comparison where no initialization has occurred. Lines are guides to the eye. (c) The average absorption of the Rabi sidebands (blue) is plotted along with the absorption in the dark state dip (green) as a function of the dark time. Clearly, NSN persists in the absence of laser illumination for well over 1s. The solid lines are an average. The black I is the error bar.}\label{Fig3}
\end{figure}

To measure the persistence of the NSN in the absence of laser interactions, we insert a dark period between the initialization and read out phases, indicated by the timing diagram in Fig. \ref{Fig3}a. The black data in Fig. \ref{Fig3}b is the spectrum with no dark period, and the red data is a comparison with no initialization or dark period. Using a pump 1 (pump 2) $\Omega_R/2\pi$ of 900MHz (650MHz) and an initialization time of 62.5ms, the absorption at both the dark state dip (green) and at the Rabi sidebands (blue) are plotted as a function of laser dark time in Fig. \ref{Fig3}c. The dark state absorption does not change, showing that the spin narrowed state persists when the lasers are shut off for a period of 1.25s (limited by the stability of the experimental apparatus) between preparation and readout. Additionally, the sample bias is being modulated according to the Stark shift modulation technique during this time. Because the modulation amplitude is large, the electron is shifted to an unstable point (co-tunneling region \cite{atature2006}) between the neutral exciton and trion bias regions \cite{warburton2000} during one half-period of the modulation cycle. The electron is randomly reinitialized at a rate of at least 3.5KHz, corresponding to the modulation frequency. This shows that NSN appears to be insensitive to the electron charge and spin orientation\cite{gong2011}. 
The hole driven nonlinear hyperfine interaction leads to a reduction of nuclear spin fluctuations (NSN) without significantly modifying the mean Overhauser field, as seen by the position of the two photon resonance. In earlier observations of extended coherence times, such effects were driven by electron spin interactions and associated with changes in the nuclear spin polarization\cite{greilich2007, bluhm2010}.

In summary, we have shown that hole-assisted dynamical nuclear polarization feedback can be used to prepare the nuclear spins in a singly charged quantum dot into a spin narrowed state which can persist in the dark for over 1 second and has a preparation time of tens of milliseconds. The spin narrowing depends only on the hole spin and appears insensitive to the electron charge and spin orientation. This means that the NSN is potentially decoupled from quantum gate operations in which detuned pulsed lasers operate only on the TPR\cite{press2008,kim2010}. Because these pulses specifically avoid populating the excited state and act primarily on the spin ground states, they should have minimal impact on the NSN. This approach can enhance the electron spin coherence prior to spin manipulation, thereby increasing the number of possible quantum computing operations without the need for spin echo coherence recovery or dynamic decoupling schemes.

\begin{acknowledgments}
We thank L.-M. Duan and Z.-X. Gong for fruitful discussions. This research was supported by the U.S. Army Research Office MURI award W911NF0910406. The authors would also like to acknowledge the NSF, AFOSR, and DARPA for their support.
\end{acknowledgments}


%

\end{document}